\documentclass[twocolumn,showpacs,prd,amsmath,amssymb,nofootinbib,balancelastpage,superscriptaddress]{revtex4}
\usepackage{psfig}
\usepackage[mathscr]{eucal}

{\catcode`\|=\active
  \gdef\Braket#1{\left<\mathcode`\|"8000\let|\bravert {#1}\right>}}
\def\bravert{\egroup\,\vrule\,\bgroup}
\newcommand{\be}{\begin{eqnarray}}
\newcommand{\ee}{\end{eqnarray}}
\newcommand{\bc}{\begin{center}}
\newcommand{\ec}{\end{center}}
\newcommand{\nn}{\nonumber}

\begin{document}

\title{Cosmic 21-cm Fluctuations as a Probe of Fundamental Physics}
\author{Matthew Kleban}
\email{mk161 [at] nyu.edu}
\affiliation{Center for Cosmology and Particle Physics, New York University, New York, NY 10003, USA}
\author{Kris Sigurdson}\thanks{Hubble Fellow}
\email{krs [at] ias.edu}
\affiliation{School of Natural Sciences, Institute for Advanced Study, Princeton, NJ 08540, USA}
\affiliation{Department of Physics and Astronomy, University of British Columbia, Vancouver, BC V6T 1Z1, Canada}
\author{Ian Swanson}
\email{swanson [at] ias.edu}
\affiliation{School of Natural Sciences, Institute for Advanced Study, Princeton, NJ 08540, USA}

%------------------------------------------------------------------------------

\begin{abstract}
Fluctuations in high-redshift 
cosmic 21-cm radiation provide a new window for observing unconventional 
effects of high-energy physics in the primordial spectrum of density perturbations.  
In scenarios for which the initial state prior to inflation is modified 
at short distances, or for which deviations from scale 
invariance arise during the course of inflation, the cosmic 21-cm power spectrum 
can in principle provide more precise measurements of exotic effects on fundamentally different scales
than corresponding observations of cosmic microwave background anisotropies.
\end{abstract}

\maketitle

%%%%%%%%%%%%%%%%%%%%%%%%%%%%%%%%%%%%%%%%%%%%%%%%%%%%%%%%%%%%%%%%%%%%%%
\section{Introduction}
\label{intro} 
%%%%%%%%%%%%%%%%%%%%%%%%%%%%%%%%%%%%%%%%%%%%%%%%%%%%%%%%%%%%%%%%%%%%%
The primary obstacle to studying fundamental physics experimentally is 
the difficulty of achieving sufficiently high energies in the laboratory.  
Conventional models of particle physics and string theory predict 
fundamentally new phenomena at energy scales of order $10^{16}$~GeV and above. 
A few observations accessible at low energies 
provide indirect clues about the underlying physics at these high 
scales, such as the long lifetime of the proton, small neutrino masses and, 
possibly, the cosmological constant.  An additional rich source of data arises from observations of large-scale structure in 
the Universe.

In the standard model of inflationary cosmology, structure is seeded by 
primordial quantum fluctuations in the inflaton
field that are stretched to super-horizon scales during $\gtrsim$ 60 e-foldings of
inflation, and subsequently collapse into the structure we observe in the 
Universe today.  During inflation, while these perturbations were forming, 
the characteristic energy scale may have been as high as 
$H_{i} \sim 10^{14}$~GeV, which is far beyond any scale accessible to laboratory experiments.  
The cosmic evolution of these small perturbations is linear, and,
if observed near the linear regime, they constitute a relatively direct window 
on physics at this extraordinarily high energy scale.  

In addition to allowing a direct probe of the physics of inflation 
(see, e.g.,~\cite{Jungman:1995bz,Lidsey:1995np,Seljak:1996gy,Kamionkowski:1996zd}), 
this observation has lead to the suggestion that the spectrum of temperature and polarization
anisotropies in the Cosmic Microwave Background (CMB) could be used as a probe 
of unconventional physics at scales at or above $H_{i}$   (see 
\cite{Brandenberger:2000wr,Martin:2000xs,Kempf:2000ac,Niemeyer:2000eh,Easther:2001fi,Kempf:2001fa,Easther:2001fz,Kaloper:2002uj,Hassan:2002qk,Danielsson:2002kx}, and references therein).  
This approach is limited by several factors:  the inflationary scale $H_{i}$, 
while high, is at best 1\% of the fiducial scale of 
$M \sim 10^{16}$~GeV, and cosmic variance, coupled with the Silk damping of anisotropies 
due to photon diffusion, limits the theoretical precision of data taken from the CMB to approximately 
that same level.  

Although its potential as a cosmological probe has long been known \cite{Field:1959,Sunyaev:1975,Hogan:1979,Scott:1990}, there has recently been renewed interest in using the 21-cm hyperfine ``spin-flip'' 
transition of neutral hydrogen as a probe of the cosmic dark ages \cite{Loeb:2003ya,Bharadwaj:2004nr,Profumo:2004qt,Sigurdson:2005cp,Sigurdson:2005mp,Zahn:2005ap,Shchekinov:2006eb,Tashiro:2006uv,Pritchard:2006sq,Furlanetto:2006wp,Cooray:2006km,Pillepich:2006fj,Metcalf:2006ji,Hirata:2006bn,Khatri:2007yv,Lewis:2007kz,Furlanetto:2006jb}.\footnote{See Ref.~\cite{Furlanetto:2006jb} for a comprehensive review of 21-cm physics, 
astrophysics, and phenomenology.}  As we will see, these 21 cm observations have the potential 
to sidestep the limitations of the CMB and provide a powerful technique 
for studying the Universe during and before inflation.  There are two reasons for this.  First, 
the data probes a different and complementary range of scales relative to the CMB (see Figure 1).  Second, 
the quantity of data available is so large (cosmic 21-cm fluctuations probe a volume rather than a surface, down to much smaller scales than CMB anisotropies) that the in-principle cosmic-variance limit on the 
precision improves dramatically (see Section \ref{21cm}).

Here, we identify two major categories of fundamental physics effects that could change the 
power spectrum of perturbations observable via cosmic 21-cm fluctuations.   The first class is {\em initial 
state effects}.  At the beginning of inflation, when the expansion of the Universe 
first began to accelerate, there is no reason to expect that the initial Hubble 
patch was close to homogeneous; instead, the details of this initial configuration 
may provide important clues for understanding the origin of the Universe, the 
nature of the big bang, and perhaps even cosmological quantum gravity.   However, 
significant periods of inflation greatly reduce the signatures of the initial state,
producing a flat and uniform Universe in the present.  The effects of a non-homogeneous 
initial state are therefore most significant shortly after the beginning of inflation, 
meaning that they are most easily visible today on very large scales.  On the other hand, it is on the 
largest scales that cosmic variance places the most significant restrictions on our 
ability to determine (even in principle) the spectrum of perturbations.  As we will see, this tradeoff 
means that certain classes of initial states are more 
easily observed (or in some cases are only observable) at the shorter scales accessible 
in 21-cm data. 

The second category addresses effects that occur 
{\em dynamically} throughout, or at some point during, the period of inflation.  These 
effects do not ``inflate away", and may occur at any time during inflation 
(and therefore affect any scale in the present).  Examples in this category include quantum 
corrections to inflaton dynamics (such as wave-function renormalizations from 
integrating out heavy fields) \cite{Kaloper:2002uj}, a field going on resonance and producing 
particles \cite{Kolb:1998ki}, a sharp feature in the inflaton 
potential \cite{Starobinsky:1992ts,Kamionkowski:1999vp}, and a myriad of other exotic 
possibilities.  Observations of cosmic 21-cm fluctuations are generally superior 
to CMB data for probing this class of effects:  they can provide much greater 
precision due to the greater amount of data available, and they can probe a 
longer ``lever arm" of data over scales inaccessible to 
CMB observations (or other probes of the matter power spectrum, such as galaxy surveys).

Our paper is structured as follows:  We first briefly review potential 
effects of high-scale physics on the spectrum of density perturbations in 
the Universe today.  We then review the physics of 21-cm fluctuations from the 
perspective of dynamics in the early Universe.  
Finally, we connect the two and discuss possible modifications to the predicted 
power spectrum of high-redshift cosmic 21-cm radiation due to new physics at high scales.

\section{High-Scale Physics and Inflation}
\label{sec:highmod}
As discussed in the introduction, we group the effects of high-scale physics into 
two categories:  initial state modifications and corrections to inflaton dynamics.

\subsection*{Initial State}
Given a scalar field evolving in a sufficiently flat potential, a region of space will begin to
inflate if it is close to homogeneous over a region roughly the size of a 
few Hubble volumes \cite{Goldwirth:1991rj}.  The state of the inflaton and any 
other relevant fields during this time constitutes the initial state for inflation.  
As the expansion proceeds, spatial gradients in these fields will stretch and 
inflate away, particles and other impurities will dilute, and the space will rapidly 
approach an approximate de Sitter phase, a state well-described by the Bunch-Davies 
vacuum \cite{Bunch:1978yq}.  Density perturbations in the initial state will be 
stretched by inflation to scales that are large in the present, so the 
effects of initial inhomogeneities will be most apparent on the largest scales.  
However, if for some reason the perturbations in the initial state had a ``blue'' 
spectrum (one with increasing power at large $k$)
with significant power on small scales (relative to the inflationary Hubble 
length), the imprint could extend down to scales significantly smaller than 
the Hubble length today.  

To make this more quantitative, we will follow the treatment of Ref.~\cite{Tanaka:2000jw}.  
We define the Bunch-Davies vacuum state $|0\rangle$ by $\hat{a}_{\bf k}|0\rangle = 0$, 
where $\hat{a}_{\bf k}$ annihilates a mode $u_{\bf k}$ 
with momentum ${\bf k}$.  An arbitrary initial state $|0\rangle_b$ in the same Fock 
space can be defined as
\be
\hat{b}_{\bf k}|0\rangle_b = 0,
\ee
where $\hat{b}_{\bf k}$ is the annihilation operator for modes $v_{\bf k}$ defined by
\be
v_{\bf k} = \alpha_{\bf k} u_{\bf k} + \beta_{\bf k} u^*_{\bf k}.
\ee
As usual, $\alpha_{\bf k}$ and $\beta_{\bf k}$ satisfy 
$|\alpha_{\bf k}|^2 = |\beta_{\bf k}|^2 + 1$.

The two-point function for the inflaton in the late-time limit is easily 
computed \cite{Tanaka:2000jw}:
\be\label{two}
\langle 0_b||\phi_{\bf k}|^2| 0_b \rangle   
	&=& {H_{i}^2 \over 2 k^3}  
	\mbox{\LARGE (}  1 + 2 |\beta_{\bf k}|^2 	
\nn
\\&&  + 2 |\beta_{\bf k}|
\sqrt{1 + |\beta_{\bf k}|^2}\cos{\varphi_{\bf k}} \mbox{\LARGE )},
\ee
where $\varphi_{\bf k} = \arg(\alpha_{\bf k}) - \arg(\beta_{\bf k})$.

This modification can naively be made arbitrarily large by 
increasing $|\beta_{\bf k}|$.  However, if the causal region is to inflate at all, 
the energy density in the perturbations must be subdominant compared to the 
vacuum energy contribution from the scalar potential.  This leads to an integrated 
constraint on the occupation numbers  $|\beta_{\bf k}|^2$ 
that must be satisfied at the beginning of inflation.  

The expectation value of the energy density (normalized with respect to the 
Bunch-Davies vacuum and time-averaged) is approximately \cite{Tanaka:2000jw}
\be\label{constraint}
\langle 0_b| \left( -T_0^0 \right) | 0_b \rangle &=& 
	{(H_{i} \eta)^4} \int \frac{d^3k}{(2 \pi)^3} 
	~k|\beta_{\bf k}|^2 , 
\\ 
\langle 0_b| \left( -T_0^0 \right) | 0_b \rangle &=& \int \frac{d^3p}{ (2 \pi)^3}\,p\,n_{\bf p} ,
\ee
where $\eta$ is the conformal time,
${\bf p}={\bf k}/a$ is the physical momentum, and the occupation number $n_{\rm p}$ is given by 
$n_{\bf p} = |\beta_{\bf k}|^2$.  We thus find that this energy density must be less than the vacuum 
energy $3 {M_{\rm Pl}^2 H_{i}^2}$ for the patch to begin inflating.\footnote{Here, $M_{\rm Pl}=\sqrt{\frac{\hbar c}{8 \pi G_{N}}}$ is the reduced Planck mass.}  However, as this is 
an integral constraint, there are a large set of initial states for which it
is satisfied, and it is apparent that some of the energy density can reside in 
short-wavelength modes.  Such initial states may be visible in the fluctuation
spectrum of cosmic 21-cm radiation.  We emphasize, however, that inflation must 
have been relatively short for these effects to be potentially visible.  
Longer inflationary periods will push the effects out to very large scales, meaning that only 
very short wavelength initial perturbations will be inside our horizon today. 
However, larger ${\bf k}$ requires smaller $|\beta_{\bf k}|$ to satisfy 
Eq.~(\ref{constraint}), implying a smaller overall effect on the spectrum [see Eq.~(\ref{two})].

If the phases in the initial state are random, the cosine 
term in Eq.~(\ref{two}) will average to 
zero, and the effect on the perturbation spectrum will be  ${\cal O}(\beta_{\bf k}^2)$.  
However, the effect can be enhanced if the phases of the ${\bf k}$-modes in the initial 
state are chosen such that $\varphi_{\bf k}$ in Eq.~(\ref{two}) are either 
constant or slowly varying 
with ${\bf k}$, in which case the effect on the perturbation spectrum will be ${\cal O}(\beta_{\bf k})$  (for $\beta_{\bf k} \ll 1$).  Such carefully chosen phases can imprint characteristic oscillation 
patterns on the perturbation spectrum (see, e.g., \cite{Danielsson:2002kx}).  

It is important to stress that, in general, $\beta_{\bf k}$ and $\varphi_{\bf k}$ are functions of the 
vector ${\bf k}$, and not just $k=|{\bf k}|$.  
In other words, because it is fixed by physical processes prior to inflation, 
the initial state need not be isotropic or random.
The potential anisotropy or structure in the initial state 
may be an important signature of the underlying 
physics, and this should be kept in mind when considering how such signatures 
arise in the (otherwise isotropic and random) power spectrum of inflationary fluctuations.

At present we have no compelling reason to claim that any particular initial 
state is preferred;  we simply aim to point out the possibility of studying the initial state
via cosmic 21 cm observations.  It is worth noting here that there are many models 
in the literature that 
predict both special initial conditions and short inflation.  In the string theoretical 
scenarios \cite{Bousso:2000xa,Susskind:2003kw}, long inflation requires fine-tuning 
beyond that required for vacuum selection.  The generic expectation is therefore 
that inflation will be relatively short, although determining precisely how short 
may be difficult.  Furthermore, if the landscape is populated by tunneling, the 
initial conditions at the big bang are quite special, and precision observations 
at large scales may even contain some information about neighboring 
minima \cite{Freivogel:2005vv,Batra:2006rz}.

\subsection*{Dynamics}
%%%%%%%%%%%%%%%%%%%%%%%%%%%%%%%%%%%%%%%%%%%%%%%%%%%%%%%%%%%%%%%%%%%%%
\begin{figure*}
\centerline{\psfig{file=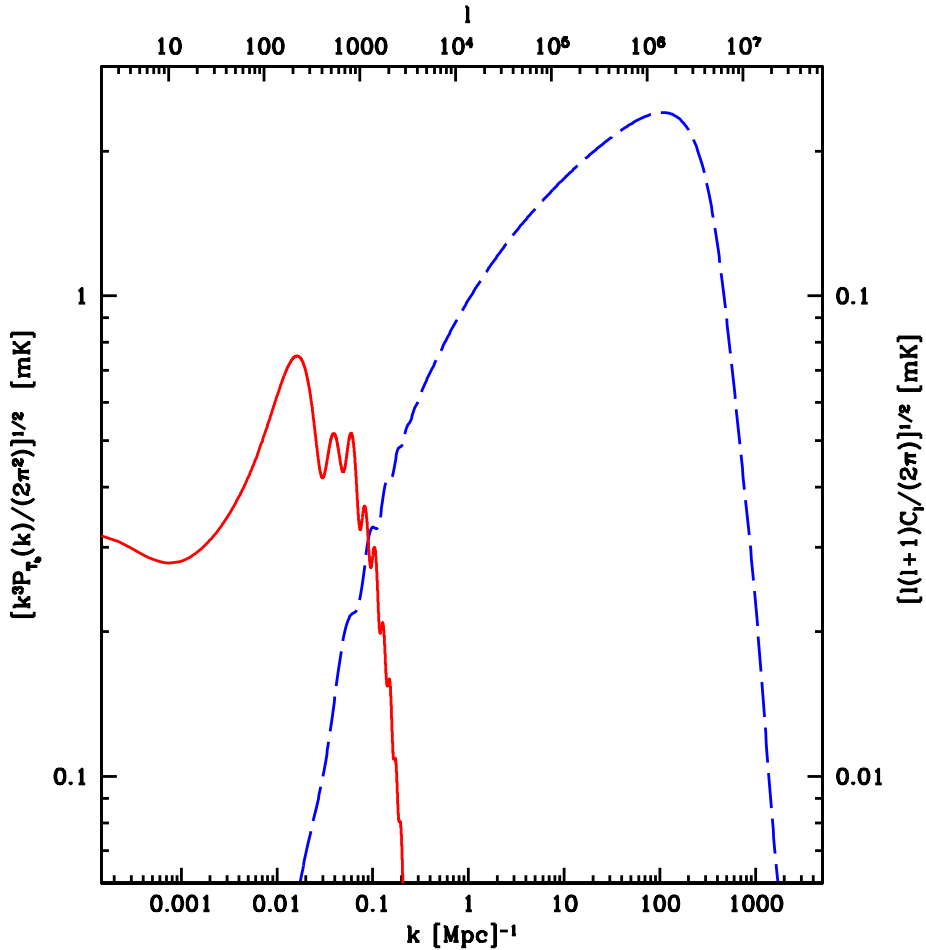,width=5.5in,angle=0}}
\caption{The scales probed by cosmic microwave background anisotropies (\emph{solid line}) 
and cosmic 21-cm fluctuations (\emph{dashed line}).  The two power spectra have been
aligned using the small-scale relation $k \simeq l/d_{\rm A}(z_{\rm CMB})$, 
where $d_{\rm A}(z_{\rm CMB})\simeq 13.6$~Gpc is the comoving angular diameter 
distance at the surface of recombination in the standard cosmological model.}
\label{fig:cmb_21}
\end{figure*}
%%%%%%%%%%%%%%%%%%%%%%%%%%%%%%%%%%%%%%%%%%%%%%%%%%%%%%%%%%%%%%%%%%%%%
There are many possible effects on inflaton physics
that can create interesting features in the spectrum of 
density perturbations during inflation.  Some examples include abrupt changes 
in the inflaton potential \cite{Starobinsky:1992ts}, and couplings to other particles that 
result in resonant production \cite{Kolb:1998ki}.  Another example involves the direct 
effect of high-scale physics on inflaton dynamics.  Such effects occur when integrating out massive fields coupled to the inflaton gives rise to wave-function renormalizations  \cite{Kaloper:2002uj}.  Because these features can occur at any time 
during inflation, the CMB alone provides a limited observational window, and is hindered 
by the constraints of cosmic variance.  Cosmic 21-cm fluctuations would therefore provide access to 
a different part of the history of inflation, and with much greater precision.  
This is illustrated in Fig.~\ref{fig:cmb_21}, where both the CMB and cosmic 21-cm power 
spectra are shown on the same figure.  While the power in the CMB anisotropies drops off 
precipitously above $l \simeq 1000$ ($k \simeq 0.07~{\rm Mpc}^{-1}$), 
the cosmic 21-cm fluctuations continue to grow, peaking near $k \simeq 100~{\rm Mpc}^{-1}$.

There are many more exotic possibilities that have appeared in the 
literature that roughly fit into this category of effects.  For instance, if the constraint 
in Eq.~(\ref{constraint}) can be avoided (for example, by using a mechanism along the lines of 
\cite{Khoury:2006fg}), there is the possibility of extending the initial state perturbations 
up to arbitrarily high frequency scales, rendering them visible throughout inflation.  
Examples of this are the de Sitter $\alpha$-vacua \cite{Mottola:1984ar,Allen:1985ux}, which, 
due to the fact they are de Sitter invariant, do not inflate away (and cannot be regarded as 
a perturbation above the Bunch-Davies state).  While we do not regard this possibility as 
plausible per se \cite{Mottola:1984ar,Kaloper:2002cs,Banks:2002nv}, we point out that 
cosmic 21-cm fluctuations could significantly constrain these and other related ideas.

%%%%%%%%%%%%%%%%%%%%%%%%%%%%%%%%%%%%%%%%%%%%%%%%%%%%%%%%%%%%%%%%%%%%%
\section{Cosmic 21-cm Fluctuations}
\label{21cm} 
%%%%%%%%%%%%%%%%%%%%%%%%%%%%%%%%%%%%%%%%%%%%%%%%%%%%%%%%%%%%%%%%%%%%%
Following the formation of the first atoms at $z \sim 1000$, the Universe became essentially 
neutral and transparent to photons.  The tiny residual population of free electrons 
was able, via Compton scattering, to couple the temperature of cosmic gas $T_g$ to the CMB 
temperature $T_{\gamma}$ until redshift $z \sim 200$.  At this point the cosmic gas began to 
cool adiabatically as $T_g \propto (1+z)^2$ relative to the CMB, which redshifts as 
$T_{\gamma} \propto (1+z)$.

During this epoch, the hyperfine spin state of the gas relevant to the 21-cm transition is 
determined by two competing processes: radiative interactions with CMB photons at 
$\lambda_{21}=21.1$~cm, and spin-changing atomic collisions.  Conventionally,
the fraction of atoms in the excited (triplet) state versus the
ground (singlet) state, 
\be
\frac{n_1}{n_0} = 3 e^{-T_{\star} / T_s} ,
\ee
is characterized by the spin temperature $T_s$.\footnote{The single spin temperature 
description outlined here is only an approximation and, in fact, the full spin-resolved 
distribution function of hydrogen atoms computed in Ref.~\cite{Hirata:2006bn} should be used 
when computing 21-cm fluctuations in detail.}  Here, $T_{\star}=hc/\lambda_{21} k_B = 68.2$~mK 
is the energy of the 21-cm transition in temperature units, and the factor of $g_1/g_0 = 3$ 
accounts for the degeneracy of the triplet state.

Spin-changing collisions efficiently couple $T_s$ to $T_g$ until $z \sim 80$, at which point 
$T_s$ rises relative to $T_g$, becoming equal to $T_\gamma$ by $z \sim 20$.  There is thus a 
window between $z \sim 200$ and $z \sim 20$ where $T_s < T_\gamma$, and fluctuations in the 
density and bulk velocity of neutral hydrogen may be seen in the absorption of redshifted 
21-cm photons.

Measurements of these fluctuations may be a powerful probe of the matter power spectrum 
\cite{Loeb:2003ya}.  The observable quantity is the brightness temperature
\be
T_b(\hat{\bf r},z) = \frac{3 \lambda_{21}^3 A T_\star}{32\pi (1+z)^2 
(\partial V_r/\partial r)}n_{H}\left(1-\frac{T_\gamma}{T_s}\right) ,
\label{eq:tb}
\ee
measured in a radial direction $\hat{\bf r}$ at redshift $z$ (corresponding to 21-cm radiation observed 
at frequency $\nu=c/[\lambda_{21}(1+z)]$).  Here, $A$ is the Einstein spontaneous emission 
coefficient for the 21-cm transition, $V_r$ is the physical velocity in the radial direction 
(including both the Hubble flow and the peculiar velocity of the gas ${\bf v}$), and  
$\partial V_r/\partial r$ is the velocity gradient in the radial direction.  Explicitly, we have
\be
\frac{\partial V_r}{\partial r} = \frac{H(z)}{1+z} + \frac{\partial 
({\bf v}\cdot \hat{\bf r})}{\partial r}  .
\label{eq:vg}
\ee
Combining Eqs.~(\ref{eq:tb}) and (\ref{eq:vg}) and expanding to linear order, we find
\be
\delta T_b=-\overline{T}_b\frac{1+z}{H(z)}\frac{\partial v_r}{\partial r}+
\frac{\partial T_b}{\partial \delta}\delta  ,
\ee
where $\overline{T}_b$ is the mean brightness temperature, $v_r={\bf v}\cdot \hat{\bf r}$ 
is the peculiar velocity in the radial direction, and 
$\delta=(n_{H}-\overline{n}_H)/\overline{n}_H$ is the overdensity of the gas.

Moving to Fourier space, we find
\begin{align}
\delta \widetilde{T}_b &=-\overline{T}_b\frac{1+z}{H(z)}\mu^2\left(ik\tilde{v}\right)+
\frac{\partial T_b}{\partial \delta}\tilde{\delta}  , \\ 
\delta \widetilde{T}_b &= \overline{T}_b\left[\mu^2+\xi\right]\tilde{\delta}  ,
\label{eq:Tbfourier}
\end{align}
where $\mu=\hat{k}\cdot\hat{r}=\cos\theta_k$ is the cosine of the angle between 
the radial direction and the direction of the wavevector ${\bf k}$, and 
$\xi$ is defined by $\xi\equiv (1/\overline{T}_b)(\partial T_b/\partial\delta)$. The second line, 
Eq.~(\ref{eq:Tbfourier}), uses the additional relation 
$\tilde{\delta} = -(ik\tilde{v})(1+z)/H$, which is, strictly speaking, valid on scales 
larger than the Jean's length during the matter dominated epoch. The total 
brightness-temperature power spectrum is thus \cite{Barkana:2004zy,Hirata:2006bn}
\be
\langle \delta\widetilde{T}_b({\bf k})\delta\widetilde{T}_b({\bf k}^{\prime})   
\rangle = (2\pi)^3\delta^{(3)}({\bf k}+{\bf k}^{\prime})P_{T_b}({\bf k})  , 
\ee
where 
\be
P_{T_b}({\bf k})= (\overline{T}_b)^2\left[\xi^2 + 2\xi\mu^2 
+ \mu^4 \right]\left|\widetilde{\Phi}(k\mu)\right|^2 P_{\delta\delta}(k)  .
\label{eq:ptb_of_mu_and_k}
\ee 
Here, $\widetilde{\Phi}(k\mu)$ is a cutoff in the magnitude of the radial wavevector 
$|{\bf k}_r|=k\mu$, and is given in detail by the Fourier transform of the 
21-cm line profile \cite{Hirata:2006bn}.  For our estimates we use a simple Gaussian 
form $\widetilde{\Phi}(k\mu)=e^{-\mu^2k^2/(2 k_{\sigma}^2)}$, with 
$k_{\sigma}\simeq500~{\rm Mpc}^{-1}$ to approximate the small-scale cutoff due to the atomic 
velocity distribution found in Ref.~\cite{Hirata:2006bn}.    After averaging over $\mu$, 
we find
\be
\!\!\!\!\!P_{T_b}({k})= (\overline{T}_b)^2\left[\xi^2{\cal E}_0(k) 
+ 2\xi{\cal E}_2(k) + {\cal E}_4(k)\right]P_{\delta\delta}(k) ,
\label{eq:ptb_of_k}
\ee 
where ${\cal E}_j \rightarrow 1/(j+1)$ as $k \rightarrow 0$, and 
${\cal E}_j \rightarrow 0$ for $k \gg k_\sigma$ (see the Appendix for further details).  
For $k < k_\sigma$, we see that the power spectrum of fluctuations in $T_b$ is 
related in a simple way to the power spectrum of fluctuations of the hydrogen density field.

%%%%%%%%%%%%%%%%%%%%%%%%%%%%%%%%%%%%%%%%%%%%%%%%%%%%%%%%%%%%%%%%%%%
\begin{figure*}
\centerline{\psfig{file=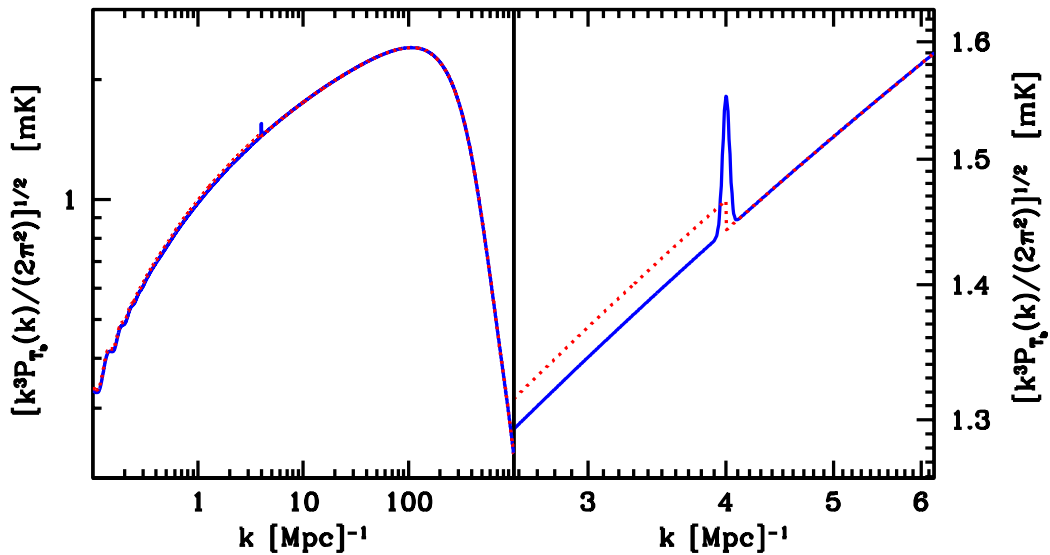,width=5.5in,angle=0}}
\caption{Left panel: 21-cm power spectra with high-energy modifications due to a remnant 
feature from the initial state of the Universe before inflation.  Shown are the effects from 
an initial state with a sharp feature centered around $k=4~{\rm Mpc}^{-1}$ (\emph{solid line}), 
and an initial state with all modes $k < 4~{\rm Mpc}^{-1}$ populated with a small 
amplitude (\emph{dotted line}).  The energy density in the initial state at the start of 
inflation is assumed to be 10$\%$ of the energy density in the inflaton potential.  For 
correlated $\varphi_{\bf k}$, these effects would occur at the level shown if 
$H_{i}/M_{\rm Pl} \sim 10^{-6}$, while for random $\varphi_{\bf k}$ they would occur 
at the level shown if $H_{i}/M_{\rm Pl} \sim 1.3 \times 10^{-7}$. Right panel:  The same 
curves enlarged to magnify the region around the spectral features. }
\label{fig:21cm_init}
\end{figure*}
%%%%%%%%%%%%%%%%%%%%%%%%%%%%%%%%%%%%%%%%%%%%%%%%%%%%%%%%%%%%%%%%%%%%%

Since primordial fluctuations in the inflaton 
field ultimately result in the fluctuations in 
baryon density and velocity that generate 
high-redshift cosmic 21-cm fluctuations, we can use the spectra of cosmic 21-cm fluctuations to study  
the types of high-scale modifications discussed above.  
To account for these effects in the present context, we apply the corrections 
to the primordial power spectrum from, for example, Eq.~(\ref{two}) 
to the power spectrum of the cosmic hydrogen field $P_{\delta\delta} (k)$.  

Before discussing this, we wish to estimate the theoretical precision of such data.
As emphasized in \cite{Loeb:2003ya}, cosmic 21 cm data correspond to a three-dimensional
volume rather than the two-dimensional last scattering surface of the CMB.\footnote{See Ref.~\cite{Kamionkowski:1997na} for another interesting method to probe our observable volume and partially circumvent cosmic variance.}  Furthermore, the 
scales involved are much smaller, implying a vastly greater number of potential data points.  
If measurements are limited by cosmic variance, the minimum relative precision using all available modes would be 
${{N}_{21}}^{-1/2} \approx (\Delta r_{\rm com}/r_{\rm com})^{-1/2}{l_{\rm max}}^{-3/2} \sim 10^{-8}$ for $\Delta r_{\rm com}/r_{\rm com} \sim 1/20$ and $l_{\rm max} \approx k_{\rm max} d_{\rm A} \sim 10^{6}$ \cite{Loeb:2003ya}.  This should
be contrasted with the corresponding cosmic variance limit for the CMB, which is roughly 
${{N}_{\rm CMB}}^{-1/2} \approx 2^{-1/2}{l_{\rm max}}^{-1} \sim 2 \times 10^{-4}$ for $l_{\rm max} \sim 3000$.  
The presence of foreground signals is unavoidable, and will reduce the number of modes 
that can be measured with a high signal-to-noise ratio.  If they are smooth in frequency-space,
however, such nuisance signals can likely be removed using techniques that will not inhibit 
measurements of the small-scale 21-cm fluctuations \cite{Wang:2005zj}.

A particularly simple type of high-scale modification of the primordial spectrum arises in 
the effective field theory that results from integrating out massive fields that
couple to the inflaton.  This will affect the power spectrum throughout inflation, so it belongs
to the second category of effects discussed in Section II.  The size of the effects 
will be ${\cal O}(H_{i}^2/M^2)$, where $M$ is the mass scale of the heavy fields 
\cite{Kaloper:2002cs}.  Due to the increased precision mentioned above, such effects are much 
easier to see (and the range of masses $M$ that can be probed is much larger) using 21-cm 
data rather than corresponding observations of the CMB.  More exotic models (which cannot be 
described by effective field theory) predict modifications of order $H_{i}/M$ 
(see, e.g., \cite{Easther:2001fi,Easther:2001fz,Easther:2002xe} and references therein), 
allowing an even greater range of $M$.  

We note that, absent an independent measurement of the inflationary Hubble scale or 
direct knowledge of the inflaton potential \cite{Kaloper:2002cs}, 
this type of effect is difficult or impossible 
to distinguish from a modification of the inflaton potential itself.  However, if 
additional information about the inflaton potential can be determined, for example, by the 
detection of inflationary gravitational waves, then it may be possible to use  
cosmic 21-cm fluctuations to probe the particle spectrum above the inflationary energy scale.

Another possibility is the observation of initial state effects.  Naively, we expect such 
effects to be most detectable on the largest scales today, and the possibility of any detection at
all assumes a short inflationary period.  However, depending on the spectrum of perturbations 
in the initial state, the effects may be more visible either in the CMB or in cosmic 21-cm 
fluctuations.  

As a simple example, consider an initial state with occupation numbers 
$n_p = \beta^2$, defined to be constant up to a physical cutoff $p_{\rm max}$, and 
$n_p = 0$ for $p > p_{\rm max}$.  For this state, the constraint in Eq.~(\ref{constraint}) 
implies that $p_{\rm max}^4 |\beta|^2 \ll 24 \pi^2 H_{i}^2 M_{\rm Pl}^2$.  For 
purposes of illustration, we demand here (and for the other initial states we consider) 
that the energy density in the initial state at the start of inflation 
(which redshifts as $a^{-4}$) is $10\%$ of the energy density in the inflaton potential.
With minimal inflation, such that $p_{\rm max}/H_{i} = k_{\rm max}/H_0$, 
and choosing $k_{\rm max} = 4~{\rm Mpc}^{-1}$, we find that 
$\beta^2 \simeq 2.7 \times 10^{-4}$ for $H_{i}/M_{\rm Pl} \sim 10^{-6}$.

If the phases of the initial state are correlated in such a way that 
$\cos \varphi_{\bf k} \sim 1$ in Eq.~(\ref{two}), 
the resulting effect on the perturbation spectrum appears at the level 
$2\beta \sim 0.03$ for all $k \lesssim 4~ {\rm Mpc^{-1}} $.  
Alternatively, if the phases of the initial state are completely random 
(such that $\cos \varphi_{\bf k} \rightarrow 0$), 
then an identical effect on the power spectrum is produced if
the Hubble scale during inflation is lowered to $\widetilde{H}_{i}$, such that 
$\widetilde{H}_{i}/M_{\rm Pl} \sim 1.3 \times 10^{-7}$ and $2 \tilde{\beta}^2 \sim 0.03$.  
Cases with equivalent and constant $|\beta_{\bf k}|$ are shown in Figure~\ref{fig:21cm_init}.  
While such an effect would be difficult or impossible to see in the CMB alone, it should be 
apparent with a clean measurement of cosmic 21-cm 
fluctuations.\footnote{Relative to this spectrum of initial 
perturbations, a blue spectrum would be easier to see using the 21 cm data, while a red spectrum 
would be more difficult.}   

Another example, also shown in Figure~\ref{fig:21cm_init}, 
is an initial state with a narrow spectral line centered at a comoving wavevector 
$k \simeq 4~{\rm Mpc}^{-1}$.  This effect is shown with $H_{i}/M_{\rm Pl} \sim 10^{-6}$,
and with correlated phases $\varphi_{\bf k}$, where the modification occurs at 
${\cal O}(\beta)$.  A similar effect can occur 
with random phases $\varphi_{\bf k}$, with $\widetilde{H}_{i}/M_{\rm Pl} \sim 1.3 \times 10^{-7}$, 
were the modification appears at ${\cal O}(\tilde{\beta}^2)$.
For simplicity, we have focused on initial states that are isotropic in ${\bf k}$ 
(depending only on $k=|{\bf k}|$), which amount to simple modifications to homogeneous 
and isotropic power spectra.  We stress that this condition can be relaxed, 
and doing so may provide an additional avenue for studying initial state effects.  
In all the cases discussed here, modifications to the standard spectrum 
of 21-cm fluctuations appear in a range inaccessible to observations of the CMB.

%%%%%%%%%%%%%%%%%%%%%%%%%%%%%%%%%%%%%%%%%%%%%%%%%%%%%%%%%%%%%%%%%%%%%
\section{Conclusions}
\label{conc}

There is a large window in momentum space
over which potential signals of fundamental high-energy (perhaps quantum-gravitational) 
effects are invisible in CMB temperature 
anisotropies, but should be apparent in the spectrum of 21-cm fluctuations.  
From the analysis presented here, one concludes
that this range runs from roughly $k \sim 0.1\, {\rm Mpc}^{-1}$, where the the CMB spectrum 
becomes strongly
suppressed, to $k \sim 1000\, {\rm Mpc}^{-1}$, 
where the cosmic 21-cm spectrum is suppressed (due to atomic velocities and the inhibition 
of growth below the Jean's length of the primordial gas).  

At the moment, the study of high-scale physics effects on the primordial
perturbation spectrum is in its infancy.  Although tantalizing hints have
emerged, we are unable to say with confidence what a generic initial state is,
or how high-scale physics might affect inflaton dynamics.
As our knowledge improves, it will be valuable to keep in mind that there may be 
effects that lie outside the regime of CMB observations, but where detection
using 21-cm observations may be possible.

%%%%%%%%%%%%%%%%%%%%%%%%%%%%%%%%%%%%%%%%%%%%%%%%%%%%%%%%%%%%%%%%%%%%%
\section*{Acknowledgments}
%%%%%%%%%%%%%%%%%%%%%%%%%%%%%%%%%%%%%%%%%%%%%%%%%%%%%%%%%%%%%%%%%%%%%
We wish to thank S.~Chang, M.~Dine, A.~Gruzinov, S.~Hellerman, 
C.~Hirata, T.~Levi, and M.~Kamionkowski for discussions.  KS thanks the Moore Center for Theoretical Cosmology and Physics at Caltech for hospitality during the final stages of this work.
KS is supported by NASA through Hubble Fellowship grant HST-HF-01191.01-A awarded by the 
Space Telescope Science Institute, which is operated by the Association of Universities 
for Research in Astronomy, Inc., for NASA, under contract NAS 5-26555. 
IS is supported by the Marvin L.~Goldberger membership at the Institute for
Advanced Study, and by US National Science Foundation grant PHY-0503584. 

\appendix
\section{The Small-Scale Cutoff}

Cosmic 21-cm fluctuations in the radial direction will have an unavoidable small-scale
cutoff due to the finite velocity distribution of hydrogen atoms \cite{Hirata:2006bn}.  
In this Appendix we calculate the anisotropy-averaged power spectrum under the 
approximation that the radial window function $\widetilde{\Phi}(k\mu)$ is a Gaussian.  

This is approximately the power spectrum that would be seen if cosmic 21-cm fluctuations 
in a redshift range $\Delta z$ (or range of comoving radius 
$\Delta r_{\rm com} = (\partial r_{\rm com}/\partial z) \Delta z$) about a central 
redshift $z_*$ were observed.  In more detailed calculations, one should use the actual radial window 
functions found in Ref.~\cite{Hirata:2006bn}, based on the non-Gaussian shape of the 21-cm line 
profile, and corrections for the evolution of 21-cm brightness-temperature 
fluctuations ($T_b(z)$, $\xi(z)$, and $\tilde{\delta} \propto 1/(1+z)$) 
across the redshift range $\Delta z$ should be included.

Starting from Eq.~(\ref{eq:ptb_of_mu_and_k}), 
we want to calculate the anisotropy-averaged power spectrum
\be
P_{T_b}(k) = \int_{-1}^{1} \frac{d\mu}{2} P_{T_b}(\bf k)  .
\ee
Defining the moments
\be
{\cal E}_{2m}(k) = \int_{-1}^{1} \frac{d\mu}{2} \mu^{2m} \left|\widetilde{\Phi}(k\mu)\right|^2,
\ee
we directly obtain Eq.~(\ref{eq:ptb_of_k}) shown above.  Specializing to the Gaussian case
\be
\widetilde{\Phi}(k\mu)=e^{-\mu^2k^2/(2 k_{\sigma}^2)} ,
\ee
(with $k_{\sigma} \simeq 500~{\rm Mpc}^{-1}$) we have
\be
{\cal E}_{2m}(k) = \int_{-1}^{1} \frac{d\mu}{2} \mu^{2m} e^{-\mu^2k^2/k_{\sigma}^2} .
\ee
In this case we have
\be
{\cal E}_{2m}(k) = \left(-\frac{k_{\sigma}^2}{2k}\frac{\partial}{\partial k}\right)^m {\cal E}_0(k) ,
\ee
where
\be
{\cal E}_0(k) = \frac{\sqrt{\pi}k_{\sigma}}{2 k}{\rm Erf}\left(\frac{k}{k_{\sigma}}\right) 
\ee
is evaluated in terms of the error function ${\rm Erf}(z)$.  
Explicitly, ${\cal E}_2(k)$ and ${\cal E}_4(k)$ appear as
\be
{\cal E}_2(k) = \frac{k_{\sigma}^2}{2 k^2}{\cal E}_0(k) 
	- \frac{k_{\sigma}^2}{2 k^2}e^{-k^2/k_{\sigma}^2}  ,
\ee
and
\be
{\cal E}_4(k) = \frac{3k_{\sigma}^4}{4 k^4}{\cal E}_0(k) - \left(\frac{3k_{\sigma}^4}{4 k^4}+\frac{k_{\sigma}^2}{2 k^2}\right) e^{-k^2/k_{\sigma}^2}  .
\ee
On large scales $k \ll k_{\sigma}$ we have
\be
{\cal E}_0(k) &\simeq& 1 - \frac{k^2}{3 k_{\sigma}^2} + \frac{k^4}{10 k_\sigma^4} + o(k^6) ,
\\
{\cal E}_2(k) &\simeq& \frac{1}{3} - \frac{k^2}{5 k_{\sigma}^2} + \frac{k^4}{14 k_\sigma^4} + o(k^6) ,
\\
{\cal E}_4(k) &\simeq& \frac{1}{5} - \frac{k^2}{7 k_{\sigma}^2} + \frac{k^4}{18 k_\sigma^4} + o(k^6) . 
\ee
\vspace{0.0001cm}

\noindent On small scales $k \gtrsim 3 k_{\sigma}$ we have
\be
{\cal E}_0(k) &\simeq& \frac{\sqrt{\pi}k_{\sigma}}{2 k}  ,
\\
{\cal E}_2(k) &\simeq& \frac{\sqrt{\pi}k_{\sigma}^3}{4 k^3}   , 
\\
{\cal E}_4(k) &\simeq& \frac{3\sqrt{\pi}k_{\sigma}^5}{8 k^5}  .
\ee

\bibliography{21cm}

\end{document}